\documentstyle[aps,prb,twocolumn,graphicx,latexsym]{revtex} 
\newcommand{\bvec}[1]{{\bf #1}}

\newcommand{\rref}[1]{Eq. \ref{#1}}

\newcommand{\qc}{q_{\rm C}}
\newcommand{\qv}{q_{\rm V}}

\newcommand{\D}[1]{D#1}

\newcommand{\beq}{\begin{equation}}
\newcommand{\eeq}{\end{equation}}

\title{Hausdorff dimension of critical fluctuations in abelian gauge 
theories}
\author{J. Hove, S. Mo, and A. Sudb{\o}}
\address{Department of Physics\\
         Norwegian University of Science and Technology,
         N-7491 Trondheim, Norway \\}
\date{\today}
\begin{document}
\bibliographystyle{plain}
\twocolumn[\hsize\textwidth\columnwidth\hsize\csname        
@twocolumnfalse\endcsname                                   
\maketitle
\begin{abstract}
  The geometric properties of the critical fluctuations in abelian
  gauge theories such as the Ginzburg-Landau model are analyzed in
  zero background field. Using a dual description, we obtain scaling
  relations between exponents of geometric and thermodynamic nature.
  In particular we connect the anomalous scaling dimension $\eta$ of
  the dual matter field to the Hausdorff dimension $D_H$ of the
  critical fluctuations, {\it which are fractal objects}. The
  connection between the values of $\eta$ and $D_H$, and the
  possibility of having a thermodynamic transition in finite
  background field, is discussed.
\end{abstract}
\draft 
\pacs{PACS numbers: 74.60.-w, 74.20.De, 74.25.Dw}
\vskip1.5pc]                                                 
Anderson has proposed the breakdown of a generalized 
rigidity associated with proliferation of defect structures in
an order parameter as a general means of characterizing phase 
transitions \cite{Anderson_deGennes}. In the context of three-dimensional 
superfluids and extreme type-II superconductors, such ideas have 
recently been put on a quantitative level \cite{tesanovic_prb99,kiet_u1_large}. 
It has been explicitly demonstrated that in three spatial dimensions abelian 
gauge theories such as the Ginzburg-Landau theory describing type-II 
superconductors, suffer a continuous phase transition driven by a proliferation 
of topological defects in the order parameter, which are closed loops of 
quantized vorticity \cite{kiet_u1_large}. These loops are induced by 
{\it transverse} phase fluctuations in a complex scalar order parameter. 
Such fluctuations are prominent in, for instance, doped Mott-Hubbard 
insulators \cite{emery_kivelson,tesanovic_prb99,kiet_u1_large}.  

In this paper, we investigate the non-trivial geometric properties of these 
critical fluctuations, and give a geometric interpretation of the anomalous 
scaling dimension of the condensate order parameter both for a charged and 
neutral condensate. In addition, we discuss the connection between the geometric 
properties of the zero-field critical fluctuations and the possibility of having 
a thermodynamic finite-field phase transition  involving unbinding of loops 
of quantized vorticity. 

We emphasize that the main results to be presented are quite general, and 
apply to the static critical sector of any theory of a complex scalar matter field 
coupled to a fluctuating gauge-field {\it in three spatial dimensions}
\cite{tesanovic_prb99,kiet_u1_large,hove1}, provided the symmetry group of the 
theory is abelian.

The Hamiltonian for the system is given by 
\begin{eqnarray}
H(q,u_{\phi}) = m_{\phi}^2 \left| \phi \right|^2 + 
\frac{u_{\phi}}{2} \left| \phi \right|^4 + 
\left| D_{\mu} \phi \right|^2 + \frac{1}{4} F^2, 
\label{glphi}
\end{eqnarray}
where $F^2 = F_{\mu \nu} F^{\mu \nu}$, 
$F_{\mu \nu} = \partial_{\mu} h_{\nu} - \partial_{\nu} h_{\mu}$,
$D_{\mu}=\partial_{\mu}-i q h_{\mu}$, and $\phi=|\phi| \exp(i \theta)$ 
is a complex matter field  coupled 
to a massless gauge field $\bvec{h}$ with coupling constant $q$. The 
$|\phi|^4$-term mediates a short-range repulsion, while the gauge-field 
${\bf h}$ mediates long range interactions. $m_{\phi}$ is the 
{\em massparameter} for the $\phi$-field, and $u_{\phi}$ is a self 
coupling. 

Consider Eq. \ref{glphi} representing a $3D$ condensate with charge 
$q \neq 0 $ sustaining stable topological objects in the form of closed
vortex loops. Then the theory with $q =0$ is a field-theoretical 
description of the ensemble of these stable topological objects, 
constituting a dual description of the original theory \cite{hove1}. 
The theory with $q=0$ is also a direct
field-theoretical description of a neutral condensate. Thus, in $3D$,
the gauge-theory $H(q\ne0,u_\phi)$ describing a {\it charged}
condensate has field-theoretical description of its critical
fluctuations or topological defects in terms of a theory isomorphic to
$H(q=0,u_\phi)$ describing a similar but {\it neutral} condensate,
and vice versa \cite{hove1}. In this sense, a charged condensate has
neutral vortices with only short-ranged steric interactions, while a
neutral condensate has charged vortices with long-ranged interactions.
In the former case, the long-ranged interactions between vortex
segments are rendered short-ranged by fluctuations of the gauge-field
in the original theory, i.e. the dual gauge-field is massive with
mass given by the charge of the original problem \cite{hove1}.  

The anomalous dimension $\eta$ for the $\phi$ field is defined via the 
relation
\beq
G({\bf x},{\bf y}) = 
\langle \phi(\bvec{x}) \phi^{\dagger}(\bvec{y}) \rangle = 
\frac{\mathcal{G} (|\bvec{x}-\bvec{y}|/\xi)}
{\left| \bvec{x} - \bvec{y} \right|^{d-2 + \eta}},
\label{corransats}
\eeq 
where ${\mathcal{G}}(z)$ is some scaling function, $\xi$ is a correlation
length, and $d$ is the spatial
dimension of the system. 
This correlation function has a geometric interpretation, yielding the
probability amplitude of finding any particle-path connecting ${\bf x}$ and 
${\bf y}$. 
In the present work, the particle 
trajectories correspond to vortex lines.

For a random walk of length $N$ in $d=3$, the probability of going from 
$\bvec{x}$ to $\bvec{y}$ is given by \cite{austin94}
\beq
P(\bvec{x},\bvec{y};N) =
\left(\frac{3}{2\pi N}\right)^{3/2}\exp\left[-\frac{\left(\bvec{x} 
- \bvec{y}\right)^2}{2N}\right].
\label{gaussiskP}
\eeq
The correlation function $G(\bvec{x},\bvec{y})$ of the corresponding
gaussian field theory is found by summing up $P(\bvec{x},\bvec{y};N)$ 
for all $N$
\beq
G(\bvec{x},\bvec{y}) = \sum_{N} P(\bvec{x},\bvec{y};N) \propto
\frac{1}{\left|\bvec{x} - \bvec{y}\right|}.
\label{gaussiskG}
\eeq
Comparing this with \rref{corransats}, we find $\eta=0$, as 
expected. The random walker traces out a fractal path with Hausdorff 
dimension $D_H$. Moreover, in general the distance between two points 
$\bvec{x}$ and $\bvec{y}$ $N$ walks apart is given by
\beq
\langle \left| \bvec{x} - \bvec{y} \right|^2 \rangle \propto N^{2\Delta},
\label{nudef}
\eeq
where $\Delta$ is the wandering exponent which for the gaussian $3D$ 
case is $\Delta = 1/2$. Inverting \rref{nudef}, we find that the total
length of the random walker scales with linear size as $L^{1/\Delta}$,
hence the Hausdorff dimension of the random walker is given by 
$D_H=1/\Delta$.  If we set $\bvec{x} = \bvec{y}$ in \rref{gaussiskP} we 
find that the unnormalized distribution $D(N)$ of {\em loops} of perimeter 
$N$, at the critical point, is given by 
\beq
D(N) \propto \frac{1}{N} \sum_{\bvec{x}}P(\bvec{0};N) \propto N^{-\alpha},
\label{gaussiskD}
\eeq 
with $\alpha=5/2$ for purely random walkers. The extra factor $N^{-1}$ 
in \rref{gaussiskD} comes from the arbitrariness in defining the starting 
position along the loop. Hence, for the case of strict random walkers in 
$3D$, described by a gaussian field theory $H=m_{\phi}^2 |\phi|^2+|\nabla \phi|^2$, 
the corresponding set of values for the two geometric and one thermodynamic 
exponents is given by $(\Delta,\alpha,\eta) = (1/2,5/2,0)$.


Beyond the gaussian case, exact exponents can not be obtained analytically, 
however we will derive scaling relations for them. When \rref{nudef} is 
invoked, a generalized probability function $P(\bvec{x},\bvec{y};N)$ may 
be written on the form
\beq 
P(\bvec{x},\bvec{y};N) \propto \frac{1}{N^{\rho}} ~
F\left(\frac{\left|\bvec{x} - \bvec{y}\right|}{N^{\Delta}}\right),
\label{generalP}
\eeq 
where $F(x)$ is a scaling function, and  normalizability of $P$ implies 
$\rho = d \Delta$. From \rref{gaussiskD}, we find that
$P(\bvec{x},\bvec{y};N)$ should scale with $N$ as $N^{1-\alpha}$,
which yields the scaling relation $\rho = \alpha-1$.
Conversely, summing over all $N$ in \rref{generalP} to find the 
correlation function $G({\bf x}, {\bf y})$, we obtain
\beq
G(\bvec{x},\bvec{y}) = \sum_{N} P(\bvec{x},\bvec{y};N) \propto
\frac{1}{\left|\bvec{x} - \bvec{y}\right|^{\frac{\rho -
        1}{\Delta} }},
\eeq
giving the scaling relation 
$\eta = \frac{\rho  - 1}{\Delta} + 2 - d$.
Combining the above, we find
\beq
\eta + \D_H = 2, ~~~ D_H = \frac{d}{\alpha-1}.
\label{combine_scaleeta}
\eeq
A computation of the {\it geometric} exponent $\alpha$ yields the 
{\it thermodynamic} exponent $\eta$ and the Hausdorff dimension 
$D_H$. Note that both $\eta$ and $D_H$ are sensitive functions of 
$\alpha$, $\partial \eta/\partial \alpha =-\partial D_H/\partial \alpha 
= d/(\alpha-1)^2$, such that a precise determination of $\eta$ 
and $D_H$ requires great precision in the determination of $\alpha$. 
The above reinforces the statement that a geometric transition of 
the vortex tangle at criticality of the gauge theory \rref{glphi} 
can be assigned a genuine thermodynamic order parameter via a dual 
formulation of the original theory {\it in three spatial dimensions}
\cite{tesanovic_prb99,kiet_u1_large}. The random walker is represented 
by a gaussian theory, Eq. \ref{glphi} with $(u_{\phi}=0,q=0)$, for which 
$\eta=0$. This corresponds to $D_H=2$, such that the random walker 
in three dimensions traces out a path that precisely fills a cross-sectional 
area of the system. {\it Note that $D_H < 2 \leftrightarrow \eta > 0$, while 
$D_H>2 \leftrightarrow \eta<0$.}

The Hamiltonian \rref{glphi} with $(u_\phi \neq 0, q \neq 0)$ has a dual 
field theory corresponding to \rref{glphi} with $q=0$ describing the neutral 
vortex tangle of a  charged superconductor \cite{hove1}. The 
$\left| \phi \right|^4$-term in \rref{glphi} represents a {\em steric 
repulsion}, i.e.  the vortex loops can not overlap, leading to a random 
walk problem with self-avoiding {\it links} (but not necessarily 
self-avoiding sites), in the sense that parallel vortex segments repel,
perpendicular vortex segments can cut \cite{cutting}, while antiparallel
vortex-segments can annihilate. Hence, this is not a standard self-avoiding 
path problem. However, we expect $\Delta > 1/2$ or equivalently $D_H<2$, 
since steric repulsion should result in a  vortex-loop tangle packing space 
less densely than for the non-interacting case, so that $\eta>0$. The 
repulsive interaction between parallel  vortex segments also leads to a 
more efficient suppression of long loops than for the non-interacting 
case, so that $\alpha > 5/2$.  

Consider next \rref{glphi} with $(u_\phi \neq0, q = 0)$ for $d=3$, which
has a dual field theory corresponding to \rref{glphi} with $q \neq 0$
describing the charged vortex tangle of a neutral condensate \cite{hove1}. 
A long-ranged (anti) Biot-Savart interaction is mediated by the gauge-field. 
This is a relevant perturbation, in renormalization group sense, to a steric 
contact repulsion\cite{dasgupta_prl_47_1981}. The geometric properties of 
the charged vortex tangle are a result of a balance between attractive
forces mediated by the gauge field, and the steric repulsion. As the numerical 
simulations show, we find $\Delta < 1/2$, corresponding to $D_H > 2$ which 
means that the vortex tangle is more compact than the ensemble of pure random 
walkers, due to the fact that an attractive long-ranged Biot-Savart interaction 
between oppositely oriented vortex segments overcompensates the steric repulsion 
so as to contract the vortex-loop tangle not only compared to the pure 
$|\phi|^4$-case, but even compared to the noninteracting case. The tangle thus 
packs space so that it more than fills a cross-sectional area of the system.

The fluctuation-dissipation theorem provides a bound on $\eta$ via the susceptibility 
$\chi_{\phi} = \int d^d x G (x) \sim \xi^{2 - \eta}$, which is bounded by the volume 
$L^d$ of the system, $L^{2 -\eta}=L^d \cdot L^{2-d-\eta}<L^d$,  so $\eta>2-d$. 
\rref{combine_scaleeta} gives a geometric interpretation of this bound. 
Specializing to $d=3$, $\eta = -1$ corresponds to topological excitations 
with $D_H = 3$, an upper limit.

For $d=3$, the continuous phase transition in a superfluid or 
extreme type-II superconductor has recently been {\it demonstrated} to be 
driven by a proliferation of vortex loops \cite{kiet_u1_large,hove1}. From 
the above, $\eta =-1$ means that a single vortex loop at $T_c$ packs space 
completely, i.e. its perimeter $N$ scales as $N \propto L^3$, implying that 
the vortex-tangle collapses on itself, rendering the transition discontinuous. 
This may be been seen from the standard scaling relation $\beta=\nu~(d-2+\eta)/2$ 
for critical exponents. Formally, this implies that the limit $\eta \to (2-d)^+$ 
corresponds to the limit $\beta \to 0^+$, characteristic of a 
discontinuous transition. More informally, a collapse of a 
vortex tangle may be viewed as mediated by an effective attractive vortex 
interaction, a situation akin to what is known in type-I superconductors. Deep 
in the type-I regime, it is known that superconductors suffer a weakly 
discontinuous transition \cite{halperin_lubensky_ma}. 

Monte Carlo simulations have been performed on the lattice
version of \rref{glphi} in the phase-only approximation, to
determine precise values of $\alpha$, both for $q=0$ and $q \neq 0$.
We have also performed simulations on pure random
walkers described by the theory $H(q = 0, u_{\phi} = 0)$. They
reveal that a determination of $\alpha$ is 
less fraught with finite-size effects
than a determination of $D_H$. Thus, we focus
on determining $\alpha$. The model we consider is
\begin{eqnarray}
H = -J \sum_{<i,j>} \cos(\theta_i - \theta_j - \qc ~ h_{i j}) 
+ \frac{1}{2} (\nabla \times \bvec{h})^2,
\label{h3dxy}
\end{eqnarray}
where the site-variable $\theta_i$ is the phase of the complex matter
field $\phi = |\phi| \exp(i\theta)$ of Eq. \ref{glphi}, when the
system is discretized, $J$ is essentially a bare phase-stiffness, 
and the link-variable $h_{ij} = \int_i^j d {\bf l} \cdot {\bf h}$. 
The charge $\qc$ is the (original) charge entering in the 
simulations. Up to this point we have considered a general charge 
$q$ irrespective of whether it couples to the original
condensate or the resulting vortex tangle. The numerical simulations
are performed on the phase of the condensate, hence the concept of
{\em original} and {\em dual} are fixed in terms of the numerical
simulations. Consequently we introduce the charges $\qc$ for the
condensate and $\qv$ for the vortices.

>From the phase distributions of the matter field we can extract vortex
loops\cite{kiet_u1_large}. These loops have charge $\qv$ and are described 
by the field theory $H(\qv,u_{\phi})$ \cite{hove1}. Hence, we can study the 
critical properties of the charged field theory $H(\qv,u_{\phi})$ by 
considering the geometric properties of the thermally excited vortex-loop 
tangle at the critical temperature in the 3DXY model. Conversely, the geometric 
properties of the vortex tangle with $\qc \neq 0$ yield the critical properties 
of the neutral field theory $H(\qv=0,u_\phi)$ of \rref{glphi}.

The simulations with $q_C=0$ are described elsewhere\cite{kiet_u1_large}, 
while for $q_C \neq 0$ the simulations proceed as follows. For every site 
on the lattice a phase change $\theta_i \to \theta_i'$ is attempted, and
accepted or rejected according to the Metropolis algorithm.
Then a change in $h_{i j} \to h_{i j} + \delta h$ is attempted, and
accepted or rejected according to the Metropolis algorithm. When
updating $h_{ij}$ we update all the link variables on a randomly
oriented elementary plaquette containing $h_{ij}$ as one of its four
edges.  Updating of $\bvec{h}$ in this fashion guarantees that the
gauge-fixing condition $\nabla \cdot \bvec{h} = 0$ is enforced at all
times.  For $\qc=0$, the simulations were performed for a system of size
$L \times L \times L$ with $L=180$, while those for $\qc \neq 0$ were
performed with $L = 64$. 

\begin{figure}[htbp]
\hfill\scalebox{0.34}{\rotatebox{-90.0}{\includegraphics{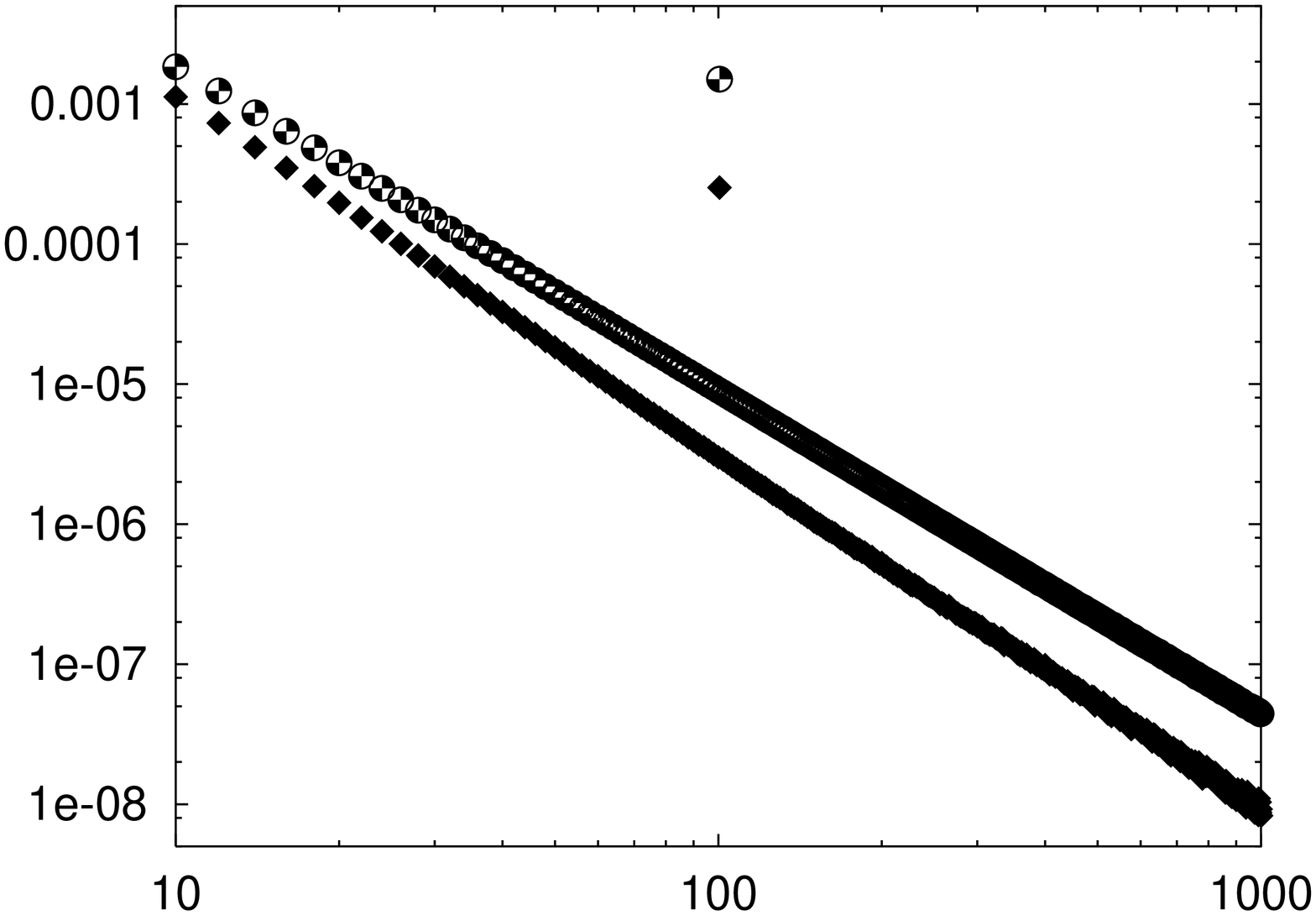}}}
\newline
\setlength{\unitlength}{1cm}
\begin{picture}(0,0)
\put(4.5  , 0.25)  {$N$}
\put(-0.1 , 3.05)   {\rotatebox{90}{$D(N)$}}
\put(5.0, 4.80)    {$\qc    = 0, \alpha = 2.31$}
\put(5.0, 5.525)   {$\qc \neq 0, \alpha = 2.56$}
\put(1.20, 3.45)   {\rotatebox{-90}{\scalebox{0.15}{\includegraphics{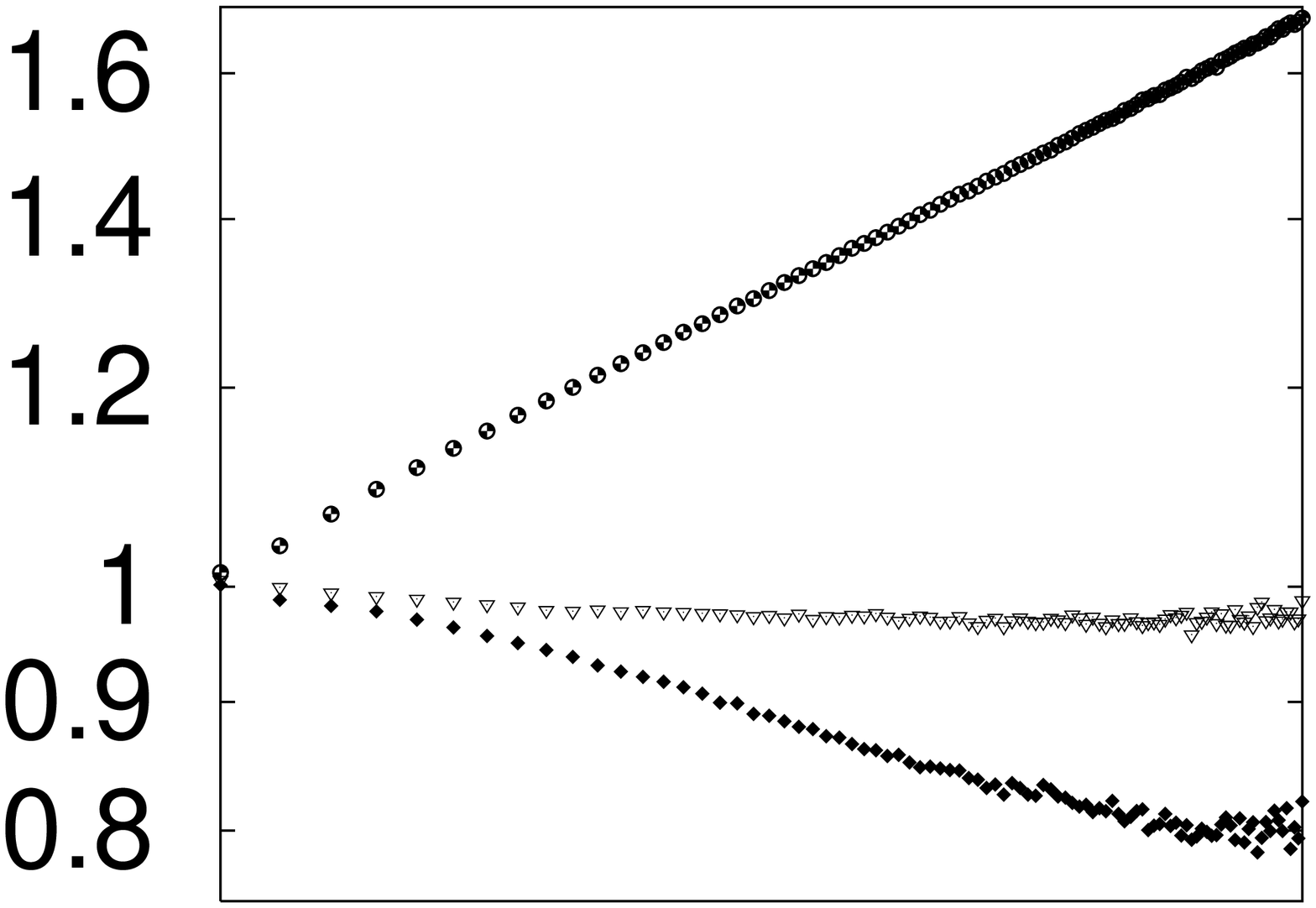}}}}
\end{picture}
\caption{\label{Dp} The vortex-loop distribution function $D(N) \sim N^{-\alpha}$ 
as a function of loop-perimeter $N$, at the critical point for the charged 
$(\qc \neq 0, \qv=0)$ and neutral $(\qc=0,\qv \neq 0)$ cases. Numerical results 
for the exponents $(\alpha,D_H,\Delta,\eta)$ are given in Table I. The system 
size is $L \times L \times L$, with $L=180$ for $q_C=0$, and $L=64$ for $q_C\neq 0$. 
Inset shows {\it {simulation results}} for $N^{5/2} D(N) \sim N^{5/2-\alpha}$ on a 
double-logarithmic scale. Top: $q_C =0,L=180$ (dual charged). Middle: Noninteracting 
vortex loops (gaussian case), $L=64$. Bottom: $q_C \neq 0,L=64$ (dual neutral). The 
results demonstrate that $5/2-\alpha > 0$ for $q_C =0$ (dual charged), while 
$5/2-\alpha < 0$ for $q_C \neq 0$ (dual neutral). Hence, by Eq. \ref{combine_scaleeta}, 
$D_H > 2, \eta < 0$ for $q_C=0$, while $D_H < 2, \eta > 0$ for $q_C \neq 0$. The 
latter agrees with other high-precision results for $\eta$, see Ref. 11. Note that 
the gaussian result $\alpha=5/2$ is obtained to high precision, for $L=64$.}
\end{figure}

During the simulations we have sampled the distribution function
$D(N)$, \rref{loop_distr}, obtaining $\alpha$. The results are 
shown in Fig. \ref{Dp} and listed in Table I. The value of $\alpha$ 
obtained for $q_C \neq 0$ (dual neutral), which is the hardest system 
to simulate, gives a value for $\eta$ in good agreement with 
high-precision results for  $\eta$ of the pure $\phi^4$-theory 
\cite{hasenbusch_99}. This serves as a useful benchmark on our method 
of extracting $\eta$. For $q_C=0$ we have simulated much larger systems 
than for $q_C \neq 0$. The deviation 
from the gaussian value $\alpha=5/2$ is substantial, {\it and of opposite 
sign compared to $q_C \neq 0$}. Given the size of the system we consider 
for $q_C=0$, it is unlikely that this is a finite-size artifact. An 
$\alpha < 5/2$ guarantees $\eta < 0$ for the $q_C=0$ (dual charged) case,
contrary to the value of $\eta > 0$ for $q_C \neq 0$
\cite{hasenbusch_99,kiet_u1_large}. In particular, the inset of Fig. 1
lends strong support to the proposition that $\eta(q_C \neq 0) > 0$, while
$\eta(q_C=0) < 0$. 

The value $\eta<0$ obtained for the original neutral, dual charged case,
is significant: It implies that $D_H>2$ for this case. {\it Whether 
$D_H > 2$ or $D_H < 2$ is of great import to the possibility of having a 
genuine phase-transition driven by a vortex-loop unbinding even in the 
presence of a finite background field such as magnetic induction in type-II 
superconductors.} A vortex system accesses configurational entropy more 
easily if it is compressible than if it is incompressible. For the 
charged case the gauge-field fluctuations render the system compressible. 
In the neutral case, the system expands screening strings of 
closed vortex loops to a larger extent than for the charged case, as substitutes 
for the gauge-field fluctuations. This is why $D_H(\qc=0) > D_H(\qc \neq 0)$. 
There is an infinitely larger amount of screening vortex-strings in the neutral 
case (dual charged) than for the charged case (dual neutral), which is the true
significance of the fact that $\eta$ is smaller for $\qc=0$ than for $\qc \neq 0$.
The possibility of the zero-field vortex-loop blowout transition surviving the 
presence of a finite field is much greater in a neutral superfluid or an extreme 
type-II superconductor, than in a charged condensate with {\it a priori} good 
screening. 

Given the significance of $D_H > 2$, we elaborate on the fact that for the 
neutral (dual charged) case, we find $\eta <0$. The Lehmann-representation of the 
Fourier transform $\tilde G(p)$ of Eq. \ref{corransats}, is sometimes used to argue 
that $\eta$ obeys the strict inequality  $\eta > 0$. The Lehmann-representation of 
$\tilde G(p)$ is given by
\begin{eqnarray}
\tilde G(k) & = & \int_{0}^{\infty} d \mu^2 ~~ 
\frac{\rho(\mu^2) }{k^2 + \mu^2}, 
\end{eqnarray}
where $1 = \int_{0}^{\infty} d \mu^2  \rho(\mu^2)$, and $\rho(\mu^2)
= Z  \delta(\mu^2-m_\phi^2) + \sigma(\mu^2)$.  The propagator for
the gaussian case would be $\tilde G(k)=1/(k^2 + m_\phi^2)$, where
$m_\phi^2$ refers to the bare massparameter in Eq. \ref{glphi}. Thus, 
$\eta > 0$ follows if $0 < Z < 1$, which holds for a uniformly positive
$\rho(\mu^2)$. However, in theories with a {\it local gauge
symmetry}, the two-point correlation function is a gauge-dependent
quantity. 
Thus, $\sigma(\mu^2)$
may in principle be made negative for certain values of $\mu$ by a
gauge-transformation. This invalidates the reasoning leading to the
strict inequalities $Z < 1$ and $\eta > 0$. 
A negative $\eta$, as found here and in other simulations
\cite{kiet_u1_large,hove1} all representing basically exact results,
agrees with a recent non-perturbative 
RG calculation \cite{Herbut}, which also gives $\nu = \nu_{3DXY}$
at the {\it charged} critical point.

At the critical point, the relevant fluctuations are {\it transverse} 
phase-fluctuations, or 
vortices \cite{tesanovic_prb99,kiet_u1_large,hove1}. Ignoring amplitude
fluctuations yields an effective  Hamiltonian governing the 
transverse $\theta$-fluctuations, whose Fourier-transform ${\cal F}$ we 
denote by ${\bf S}_k$, ${\cal F}((\nabla \theta)_T)={\bf S}_k=
-2 \pi ~ i ~ ({\bf k} \times {\bf n}_k)/k^2$, where ${\bf n}_k$ is the 
Fourier-transform of the {\it local vorticity}. We find, after integrating 
out the transverse gauge-field, that $H=\Xi^2(k)~~{\bf S}_k \cdot {\bf S}_{-k}$, 
where $\Xi^2(k)=k^2/(k^2+2q^2)$. 
For $q=0$, we have $\Xi^2(k)=1$, while 
$\lim_{k\to 0} \Xi^2(k) \sim  k^2$ for $q \neq 0$. The coupling to a 
fluctuating gauge field {\em softens} the transverse phase-fluctuations,
providing the effective phase-stiffness with an extra power $k^2$ compared 
to the $q=0$-case. Thus, $\tilde G^{-1}(k,q=0)=k^2+\Sigma(k)$ and 
$\tilde G^{-1}(k,q \neq 0)=k^4+\Sigma(k)$. In both cases, the $k \to 0$-limit 
of the self-energy $\Sigma(k)$ is given by $\Sigma(k) \sim k^{2-\eta}$. We 
thus have $\lim_{k \to 0} \tilde G^{-1}(k) \sim k^{2-\eta}$ provided 
$\eta > 0$ for $q=0$ and, when invoking the absolute lower bound, $\eta>-1$ 
for $q \neq 0$. For a pure $|\phi|^4$-theory,  
the Lehmann-representation coupled with positive-definiteness of $\rho(\mu^2)$, 
holds.

This work was supported by the Norwegian Research Council via the High 
Performance Computing Program, and by Grant 124106/410 (S.M. and A.S). 
(J. H.) thanks NTNU for support. Communications with F. S. Nogueira and 
Z. Te{\v s}anovi{\'c}, are gratefully acknowledged.


\begin{table}[htbp] 
\begin{center}
  \begin{tabular}{c|c|c|c} \hline
    Exponent  & Gaussian & $\qc=0,\qv \neq 0$ & $ \qc \neq 0,\qv=0$ \\ \hline
    $\alpha$  &5/2 & 2.312 $\pm$ 0.003  &2.56 $\pm$ 0.03   \\ \hline 
    $D_H$     &2   & 2.287 $\pm$ 0.004  &1.92 $\pm$ 0.04   \\ 
    $\Delta$  &1/2 & 0.437 $\pm$ 0.001  &0.52 $\pm$ 0.01   \\
    $\eta$    &0   &-0.287 $\pm$ 0.004  &0.08 $\pm$ 0.04   \\
  \end{tabular}
\caption{\label{tabella} The loop distribution exponent $\alpha$, as
    determined from Monte-Carlo simulations. The remaining exponents 
    have been determined from Eq. \ref{combine_scaleeta}. Symbols are 
    explained in the text. }
\end{center}
\end{table}
\end{document}